\documentclass[onecolumn,draftclsnofoot,12pt]{IEEEtran}
\usepackage{amssymb,amsmath}
\usepackage[dvips]{graphicx}
\usepackage{amsfonts}
\usepackage{subfigure}
\usepackage{booktabs,multirow}
\usepackage{color}
\usepackage{cite}
\usepackage[table]{xcolor}
\usepackage{enumerate}

\IEEEoverridecommandlockouts

\begin{document}

\title{5G Software Defined Vehicular Networks}

\author{\normalsize
Xiaohu Ge$^1$,~\IEEEmembership{Senior~Member,~IEEE,} Zipeng Li$^1$,~\IEEEmembership{Student~Member,~IEEE,} Shikuan Li$^1$\\
\vspace{0.70cm}
\small{
$^1$School of Electronic Information and Communications,\\
Huazhong University of Science and Technology, Wuhan 430074, Hubei, P. R. China.\\

Email: \{xhge, zipengli91, m201671826\}@mail.hust.edu.cn}\\

\thanks{\small{ Submitted to IEEE Communications Magazine on SI of Software Defined Vehicular Networks: Architectures, Algorithms and Applications.}}
\thanks{\small{The authors
would like to acknowledge the support from the NSFC
Major International Joint Research Project (Grant No. 61210002), the Fundamental Research Funds for the Central Universities under the grant 2015XJGH011. This research is partially supported by the EU FP7-PEOPLE-IRSES, project acronym WiNDOW (grant no. 318992) and project acronym CROWN (grant no. 610524), China International Joint Research Center of Green Communications and Networking (No. 2015B01008). }}
}

\renewcommand{\baselinestretch}{1.2}
\thispagestyle{empty}
\maketitle
\thispagestyle{empty}

\setcounter{page}{1}\begin{abstract}
With the emerging of the fifth generation (5G) mobile communication systems and software defined networks, not only the performance of vehicular networks could be improved but also new applications of vehicular networks are required by future vehicles, e.g., pilotless vehicles. To meet requirements from intelligent transportation systems, a new vehicular network architecture integrated with 5G mobile communication technologies and software defined network is proposed in this paper. Moreover, fog cells have been proposed to flexibly cover vehicles and avoid frequently handover between vehicles and road side units (RSUs). Based on the proposed 5G software defined vehicular networks, the transmission delay and throughput are analyzed and compared. Simulation results indicate that there exist a minimum transmission delay of 5G software defined vehicular networks considering different vehicle densities. Moreover, the throughput of fog cells in 5G software defined vehicular networks is better than the throughput of traditional transportation management systems.

\end{abstract}

\IEEEpeerreviewmaketitle

\newpage
\section{Introduction}
Nowadays the fifth generation (5G) mobile communication systems are developed by industrial and academic researchers. With the development of millimeter wave and massive multi-input multi-output (MIMO) technologies, the spectrum efficiency and energy efficiency are obviously improved for 5G wireless communications \cite{1Chen,2Gong}. With the emergence of pilotless vehicles, some rigorous requirements, e.g., the transmission delay need to be less than 1 millisecond, are required for intelligent transportation systems (ITSs) and vehicular networks \cite{3Ge}. To meet these rigorous requirements, 5G mobile communication technologies, cloud computing and software defined networks (SDNs) are expected to be integrated into future vehicular networks. Therefore, it is necessary to design a new network architecture for 5G vehicular networks.

Some basic issues have been investigated for vehicular networks \cite{4Araniti,5Karagiannis,6Zhang,7Taleb}. Considering drawbacks of IEEE 802.11p networks, e.g., the poor scalability, the low capacity and the intermittent connectivity, the long term evolution (LTE) mobile communication technologies were proposed to support vehicular applications \cite{4Araniti}. Moreover, the open issues of LTE vehicular networks were discussed to promote potential solutions for future vehicular networks. In reference \cite{5Karagiannis} the basic characteristics of vehicular networks was introduced. An overview of applications and associated requirements was presented and challenges were discussed. Besides, the past major ITS programs and projects in United States of America (USA), Japan and Europe were analyzed and compared. An analytical model supporting multi-hop relay of infrastructure-based vehicular networks was proposed to analyze uplink and downlink connectivity probabilities \cite{6Zhang}. Simulation and experiments results revealed that there exists a trade-off between the proposed performance metrics and system parameters, such as base station (BS) and vehicle densities, radio coverage and the maximum number of hops in a path contains. When the LTE communication technologies have been integrated into vehicular networks, the interference cuts down the performance of LTE vehicular networks \cite{7Taleb}. To overcome this issue, the millimeter wave transmission technology was proposed to connect users inside vehicles. On the other hand, the SDN was proposed as an effective network technology to be capable of supporting the dynamic nature of vehicular network functions and intelligent applications while lowering operation costs through simplified hardwares, softwares and managements \cite{8Sezer}. Consequently, some initial studies have been carried out to integrate the SDN technology into vehicular networks \cite{9Jutila,10Liu}. Utilizing the SDN technology, an adaptive edge computing solution based on regressive admission control and fuzzy weighted queueing was proposed to monitor and react to network QoS changes within vehicular network scenarios \cite{9Jutila}. Based on the SDN technology, a cooperative data scheduling algorithm integrated at road side units (RSUs) was developed to enhance the data dissemination performance by exploiting the synergy between infrastructure-to-vehicle (I2V) and vehicle-to-vehicle (V2V) communications \cite{10Liu}. However, the SDN technology in reference \cite{10Liu} is limited in RSUs. When a lot of vehicles are connected with a RSU, the frequent handover problem will reduce the performance of SDN at RSUs of vehicular networks \cite{11TTaleb}.

To meet the high performance requirements, such as the low transmission delay and high throughput, a new architecture of 5G software defined vehicular network is proposed in this paper. The main contributions of the proposed 5G software defined vehicular network are list as follows:

\begin{enumerate}[1)]
\item Based on the basic functions and requirements of vehicular networks, an architecture of 5G software defined vehicular network integrated with the SDN, cloud computing and fog computing technologies is proposed to form three logistical planes in network architecture, i.e., the application plane, the control plane and the data plane. Based on three logistical planes of network architecture, the control and data functions of 5G software defined vehicular networks are separated to improve the flexibility and scalability of vehicular networks.
\item The fog cell structure is proposed and performed at the edge of 5G software defined vehicular networks. Based on the fog cell structure, the frequent handover between the RSU and vehicles is avoided and the adaptive bandwidth allocation scheme is adopted for vehicles in fog cells.
\item The transmission delay and throughput of 5G software defined vehicular networks are analyzed. Simulation results indicate that there exist a minimum transmission delay of 5G software defined vehicular networks considering different vehicle densities. Moreover, the throughput of fog cells in 5G software defined vehicular networks is better than the throughput of traditional transportation management systems.
\end{enumerate}

In this article we propose a new architecture of 5G software defined vehicular networks adapting the cloud computing and fog computing technologies. Moreover, the control plane and data plane are separated by the SDN technology in 5G software defined vehicular networks. To avoid frequent handover between the RSU and vehicles, the fog cell is structured and multi-hop relay method is adopted for vehicular communications in a fog cell. Furthermore, the transmission delay and the throughput of fog cells are simulated for 5G software defined vehicular networks. Finally, the challenges of vehicular networks are discussed and the conclusion is drawn.

\section{5G Software Defined Vehicular Networks}

\subsection{Topology Structure of 5G Software Defined Vehicular Networks}
The cloud computing and fog computing technologies are emerging for applications of 5G vehicular networks. Moreover, the SDN is becoming a flexible approach to connect wireless access networks and clouding computing centers for 5G vehicular networks. Based on cloud computing and fog computing technologies, a 5G software defined vehicular network is proposed in this paper. The topology structure of 5G software defined vehicular networks is illustrated by Fig. 1(a). 5G software defined vehicular networks are composed of the cloud computing centers, SDN controllers (SDNCs), road side unit centers (RSUCs), road side units (RSUs), BSs, fog computing clusters, vehicles and users. Moreover, 5G software defined vehicular networks include infrastructure to infrastructure (I2I) links, vehicle to infrastructure (V2I) links and vehicle to vehicle (V2V) links. Based on 5G software defined vehicular networks, the information is shared among vehicles and users under the control of the fog computing clusters. To support the promptly responses from vehicles and users, fog computing clusters are configured in the edge of 5G software defined vehicular networks. The network structure of fog computing clusters is a distributed networks. Most of data in the edge of 5G software defined vehicular networks is saved and processed by fog computing clusters which include the RSUC, RSUs, BSs, vehicles and users. The SDNCs collect and forward the state information of fog computing clusters into the cloud computing centers. Moreover, the control information is replied to fog computing clusters by SDNCs. The core of 5G software defined vehicular networks composed by SDNCs and cloud computing centers is adopted by a centered network structure which focuses on the data forwarding and resource allocation. The detail logical structure of 5G software defined vehicular networks is described in Fig. 1(b).

\begin{figure}[!t]
\centering
\subfigure{\includegraphics[width=5.5in]{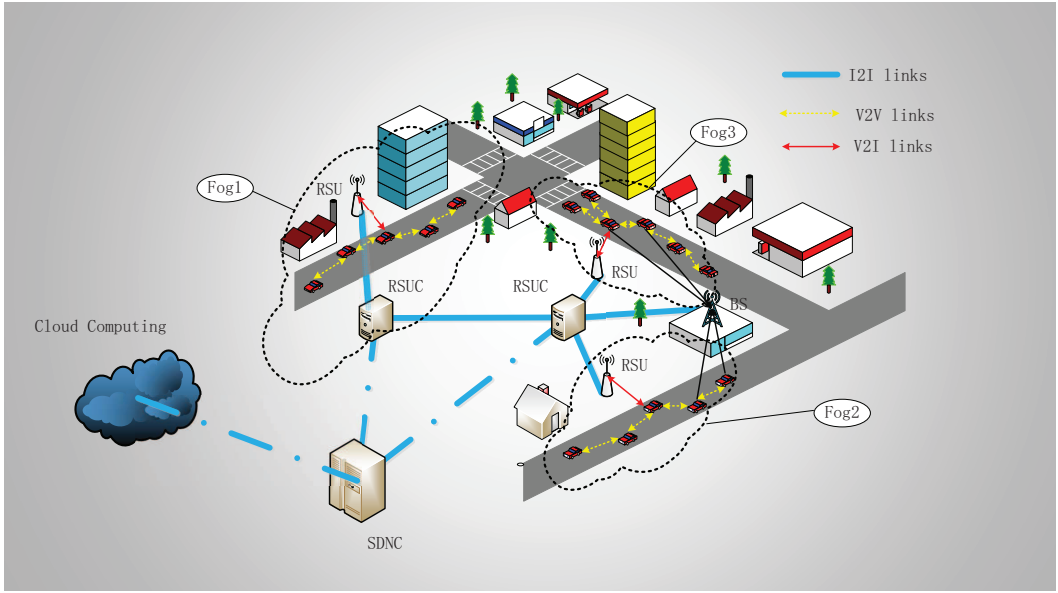}}\\
(a)\\
\subfigure{\includegraphics[width=4.5in]{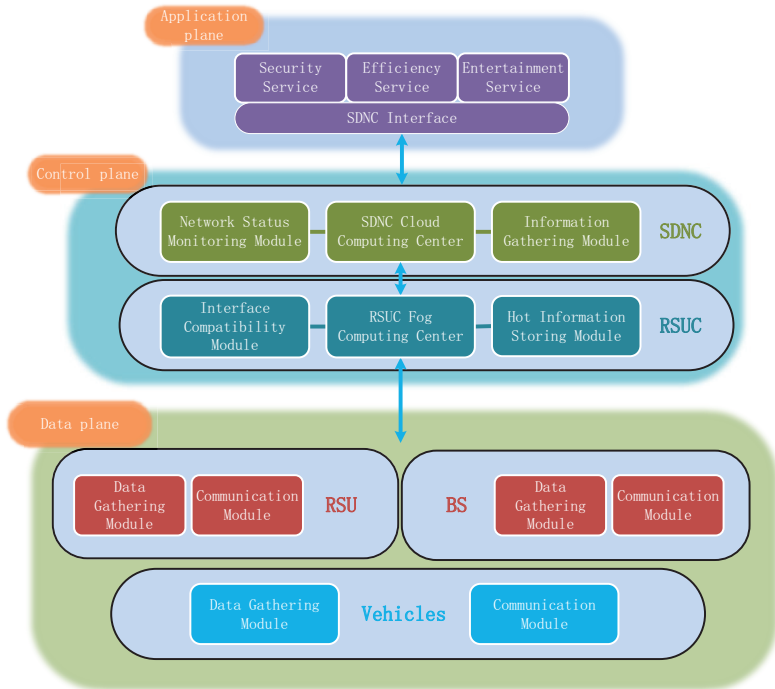}}\\
(b)\\
\caption{(a) Topology structure of 5G software defined vehicular networks; (b) Logical structure of 5G software defined vehicular networks}\label{fig1}
\end{figure}

\subsection{Logical Structure of 5G Software Defined Vehicular Networks}
In Fig. 1(b), the Logical structure of 5G software defined vehicular networks is composed by the data plane, the control plane and the application plane.

\begin{enumerate}[a)]
\item The data plane includes vehicles, BSs and RSUs. Functions of the data plane are focused on the data collection, quantization and then forwarding data into the control plane \cite{12Feng}. In detail, the vehicle can be configured with the following function modules:

\begin{itemize}
  \item
     Information collection module of vehicles: the information collection module is made up of different types of sensors in the vehicle. Utilizing sensors in the vehicle, the information of the vehicle (e.g., the speed, direction and type of the vehicle) and the environment (e.g., the number of adjacent vehicles, the users in the vehicle and the road under the vehicle) is collected for 5G software defined vehicular networks.
  \item
     Position information module of vehicle: The position information of vehicles include the independent position information and the dependent position information. In general, the independent position information of vehicles is obtained by the global position systems (GPS) which provide the detail location of vehicles at the longitude and latitude of the earth. The dependent position information of vehicles is obtained by sensors of vehicles which provide the distance between adjacent vehicles. Compared with the independent position information of vehicles, the dependent position information of vehicles can provide a high location precision for 5G software defined vehicular networks.
  \item
      Communications module of vehicle: The communication module includes V2I and V2V communication modules. The V2I communication module provides the wireless communication between vehicles and the infrastructure along the road. The V2V communication module provides the wireless communication among adjacent vehicles.
\end{itemize}
\end{enumerate}

BSs can provide the wireless communication for vehicles and RSUC. In 5G software defined vehicular networks, BSs transmit wireless signals by traditional long term evolution (LTE) frequency and provide a large coverage for vehicles. In general, vehicles first access with RUSs but secondly access with BSs when RUSs can not provide the enough resource for wireless accessing in 5G software defined vehicular networks.

In 5G software defined vehicular networks, RSUs can be configured with the following function modules:

\begin{itemize}
  \item
     Information collection module of RSU: Composed by different sensors, e.g., cameras and speed measurement sensors. The information collection module of RSUs can provide the speed of vehicles, the traffic status and the road status, etc.
  \item
     Communication module of RSU: Including two types of links, one is the links between RSUs and the RSUC and the other is the links between RSUs and vehicles. The links between RSUs and the RSUC are performed by fronthaul links in 5G software defined vehicular networks.
\end{itemize}

\begin{enumerate}[b)]
\item The control plane includes RSUCs and SDNC. The RSUC is the control center of a fog cell. Considering the quick mobility of vehicles and the massive wireless traffic between the RSU and vehicles, the frequent handover should be avoided for wireless communications between the RSU and vehicles. To solve this issue, the fog cell is proposed for 5G software defined vehicular networks. A fog cell is composed by vehicles and a RSU. Millimeter wave links are adopted for wireless relay communications among vehicles and the total bandwidth of millimeter wave is shared by all vehicles in a fog cell. Since all vehicles orderly move in an urban road, the total vehicle group can be assumed to be an overall communication unit within millimeter wave links in a fog cell. When one of vehicles in a vehicle group connects with the RSU, the total vehicle group in the fog cell could be connected with the RSU. In this case, the frequent handover can be avoided for vehicles and the RSU in a fog cell. Hence, the RSUC is configured to allocate resources and improve the transmission efficiency in a fog cell. The SDNC is the total control center for 5G software defined vehicular networks and allocates resources among fog cells. Therefore, the control plane takes charge of drawing the global information map based on the data information forwarded from the data plane and then generating the control information based on rules and strategies from the application plane. To support above functions of the control plane, RSUCs and SDNC are configured with the following function modules:

\begin{itemize}
  \item
     Information collection modules of RSUC and SDNC: drawing the global information map based on the data information from the data plane.
  \item
     Networking status module: monitoring the link status of 5G software defined vehicular networks \cite{13Bradai}.
  \item
     Computing module: deriving the control results based on the global information map and the link status of 5G software defined vehicular networks. In general, computing modules are deployed at the cloud computing center and fog computing centers \cite{14Ge}.
  \item
     Hot caching module: saving the popular data context at RSUCs to decrease the transmission delay for vehicle applications.
\end{itemize}
\end{enumerate}

\begin{enumerate}[c)]
\item The application plane directly faces different application requirements from users and vehicles. Based on application requirements from users and vehicles, rules and strategies of 5G software defined vehicular networks are generated by the application plane and forwarded into the control plane. In general, the application plane includes the security service module, the service efficiency module and the entertainment service module.
\end{enumerate}

Based on the logical structure of 5G software defined vehicular networks in Fig. 1(b), the data plane takes charge of collecting data, the control plane takes charge of deriving control instructions and the application plane takes charge of generating rules and strategies.

\section{Transmission delay and Throughput of 5G Software Defined Vehicular Networks}

Without loss of generality, the transmission delay and throughput analysis are investigated in a fog cell of 5G software defined vehicular networks. A typical fog cell is composed by a RSU and a number of vehicles in Fig. 2. To avoid frequent handover between the RSU and vehicles in the fog cell, a vehicle, i.e., the gateway vehicle is selected to connect with the RSU and then other vehicles are connected with the gateway vehicle by a multi-hop relay method. When a gateway vehicle is located in the coverage region of the RSU, the gateway vehicle directly communicates with the RSU. When other vehicles are located in the fog cell, even these vehicles do not directly coveraged by the RSU in the fog cell, these vehicles will build a multi-hop relay route to connect with the gateway vehicle and then the gateway vehicle will forward those requests/data with the RSU in the fog cell. When the gateway vehicle departs from the fog cell, a vehicle in the fog cell is handed off to service as the gateway vehicle \cite{15Stojmenovic}. In this way, all vehicles in the fog cell can maintain wireless communications with RSU while moving along the road. Since the fog cell is the basic composition of proposed 5G software defined vehicular networks, the transmission delay and throughput of the vehicle in a fog cell is investigated in the following sections.

\begin{figure}[!t]
\begin{center}
\includegraphics[width=6.5in]{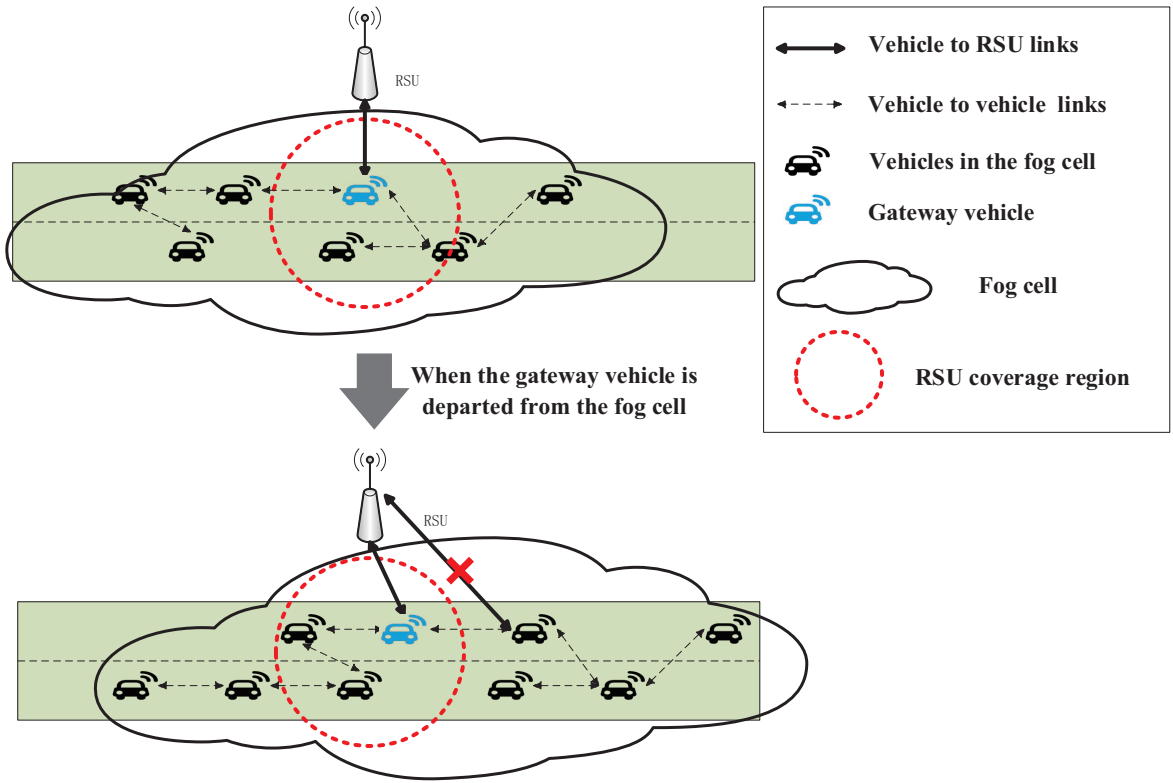}
\caption{Vehicle communications in a typical fog cell}\label{Fig2}
\end{center}
\end{figure}

\subsection{Transmission Delay of 5G Software Defined Vehicular Networks}
The transmission delay is one of core metrics for 5G software defined vehicular networks. In this paper, the transmission delay of the vehicle in a fog cell is analyzed for 5G software defined vehicular networks.

\begin{figure}[!t]
\begin{center}
\includegraphics[width=6.5in]{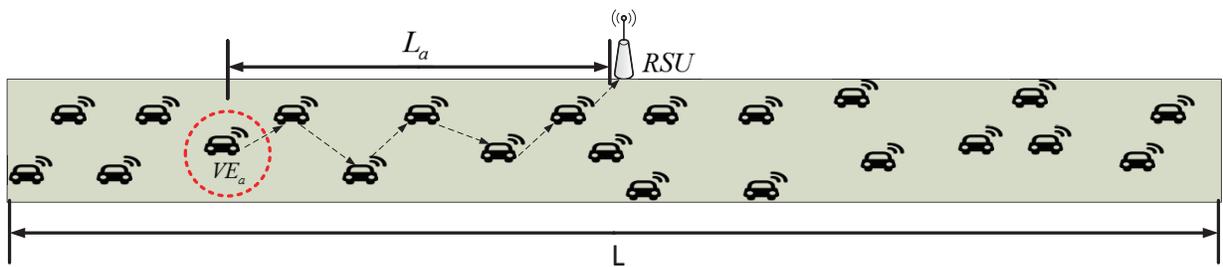}
\caption{Transmission delay in a fog cell of 5G software defined vehicular networks.}\label{Fig3}
\end{center}
\end{figure}

In Fig. 3, a RSU is located at a fog cell to service for all vehicles driving at a road with the length of $L$. Without loss of generality, a vehicle marked with red dashed circle, i.e., $V{E_a}$ is selected to analyze the transmission delay in a fog cell of 5G software defined vehicular networks. The distance between the RSU and the vehicle $V{E_a}$ is denoted as ${L_a}$. Based on the vehicle communication scheme in Fig. 2, the data packet generated from $V{E_a}$ is transmitted to the RSU by a multi-hop vehicle relay method.

Assumed that there exist $k$ hops between the RSU and the vehicle $V{E_a}$. For a data packet, the transmission delay in a fog cell of 5G software defined vehicular networks is expressed as $T = k{T_{hop}} + \left( {k - 1} \right){T_{retran}}$ , where ${T_{hop}}$ is the average transmission delay in one hop of vehicle communications, ${T_{retran}}$ is the retransmission delay which is the relay processing time at the relay vehicles. In a hop of vehicle communications, the wireless transmission is time slotted and one data packet is transmitted in each time slot ${t_{slot}}$. Assumed that the success transmission probability of vehicle relay communications is ${P_{hop}}$. As a consequence, the average transmission delay in one hop of vehicle communications is calculated by ${T_{hop}}{\rm{ = }}\frac{{{t_{slot}}}}{{{P_{hop}}}}$.

In this paper the millimeter wave transmission technique is adopted for vehicle relay communications. Without loss of generality, the 60 GHz frequency spectrum is assumed to be used for vehicle relay communications. Since the wireless signal of vehicle relay communications are usually transmitted in line of sight (LOS) scenarios, the interference is ignored for the vehicle relay communications in this paper. When the signal-to-noise ratio (SNR) threshold at the receiver is assumed as $\theta $, i.e., the data packet can be successfully received only if the SNR of receive signal is larger than the threshold $\theta $, the success transmission probability ${P_{hop}}$ is calculated by ${P_{hop}} = {\rm P}\left( {PL \le {P_{tx}}(dB) - \theta (dB) - {N_0}{W_{mmWave}}(dB)} \right)$, where $PL[dB](\delta ) = 69.6 + 20.9\log (\delta ) + \xi$, ${\rm{ }}\xi  \sim \left( {0,{\sigma ^2}} \right)$ is the path loss fading over millimeter wave wireless channels, $\delta $ is the wireless transmission distance between the transmitter and receiver, ${P_{tx}}$ is the transmission power of vehicles, ${N_0}$ is the noise power spectrum density, ${W_{mmWave}}$ is the bandwidth of millimeter wave links.

To analyze the transmission delay in a fog cell of 5G software defined vehicular networks, the default parameters are configured as follows: the noise power spectrum density is ${N_0} =  - 174$ dBm/Hz, the bandwidth of millimeter wave links is ${W_{mmWave}} = 2$ GHz, the retransmission delay is ${T_{retran}} = 5$ microsecond, one time slot is ${t_{slot}} = 5$ microsecond. Moreover, the transmission distance of millimeter wave communications is limited into 50 meters.

\begin{figure}[!t]
\begin{center}
\includegraphics[width=6.5in]{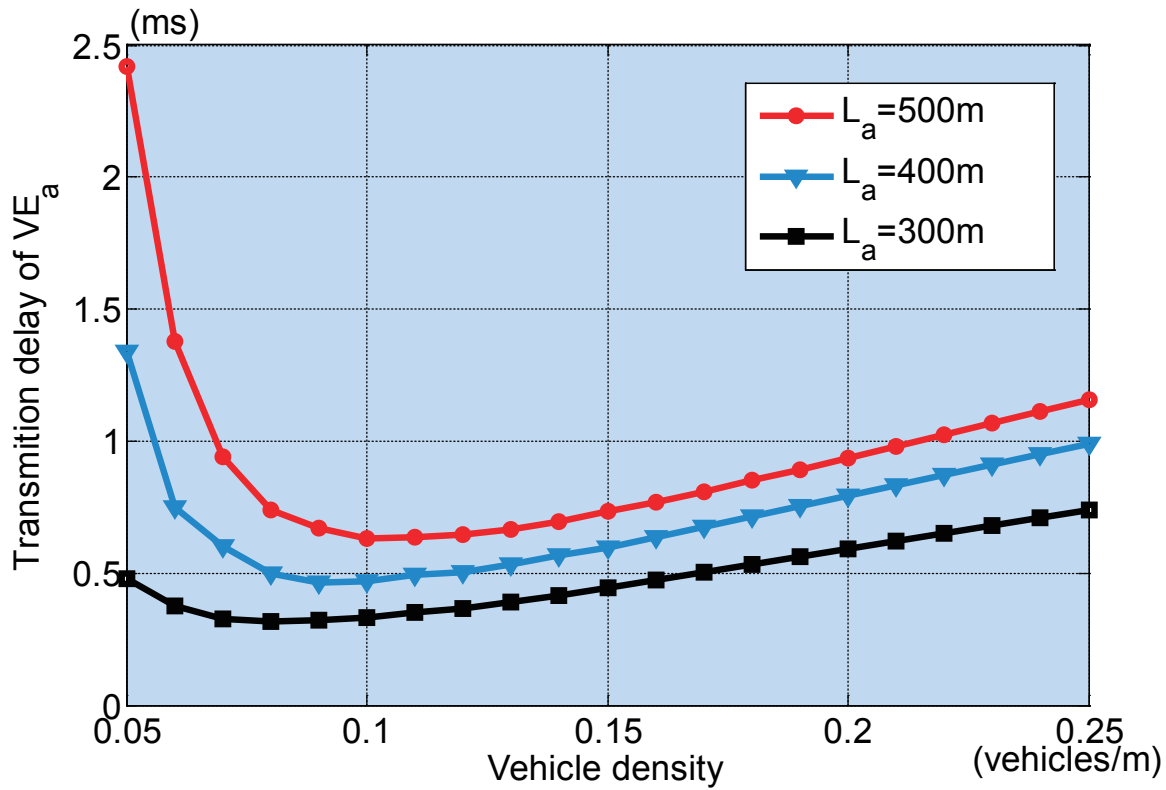}
\caption{Transmission delay with respect to the vehicle density considering different transmission distances}\label{Fig4}
\end{center}
\end{figure}

Fig. 4 illustrates the transmission delay in a fog cell of 5G software defined vehicular networks with respect to the vehicle density considering different transmission distances ${L_a}$. When the vehicle density is fixed, the transmission delay increases with the increase of the transmission distances ${L_a}$. When the transmission distance ${L_a}$ is fixed, the transmission delay first decreases with the increase of the vehicle density. However, numerical results indicate there exist turning points for vehicle densities ( the turning points are 0.08, 0.09 and 0.105 for ${L_a} = 300$ , 400 and 500, respectively). When the vehicle density larger than or equal to the turning point, the transmission delay increases with the increase of vehicle density.

The numerical results in Fig. 4 show that there exists a minimum value for the transmission delay in the fog cell of 5G software defined vehicular networks. The minimum transmission delay are 0.32, 0.46 and 0.63, corresponding to the transmission distance of 300, 400 and 500 meters, respectively. When the vehicle density is low, the distance among adjacent vehicles is far away and then the success transmission probability of millimeter wave links is low. In this case, the increasing of vehicle density will decrease the distance among adjacent vehicles and then increase the success transmission probability of millimeter wave links. Hence, the transmission delay first decreases with the increase of the vehicle density. When the vehicle density is larger than a threshold, the distance among adjacent vehicles is closed and the success transmission probability of millimeter wave links approaches to a stationary value. In this case, the transmission delay mainly dependent on the retransmission delay in every hop vehicle communication. In this case, the increasing of vehicle density will increase the number of relay hops and then the total retransmission delay is increased. Therefore, the transmission delay first increases with the increase of the vehicle density.

\subsection{Throughput of 5G Software Defined Vehicular Networks}

In traditional bandwidth allocation scheme, all bandwidths are averagely allocated to every vehicle in a fog cell. However, every vehicle needs different bandwidth in practical applications. Based on the control function of 5G software defined vehicular networks which is realized at the RSUC, an adaptive bandwidth allocation scheme is proposed to optimize the throughput of fog cells in this paper.

Without loss of generality, the available bandwidth in a fog cell is assumed as $B$ and the maximum throughput of this fog cell is $C$. The average bandwidth requirement of one vehicle is ${B_{ave}}$ and the throughput of this vehicle is configured as ${C_{ave}}$. The total number of vehicles in a fog cell is assumed as $N(N > 0)$ and the signal-to-noise ratio (SNR) at every vehicle is configured as the same. The interference is ignored in this paper. Hence, in this paper the throughput of vehicle is proportion to the communication bandwidth of vehicle. When there are $N$ vehicles in the fog cell, the bandwidth requirement of vehicles is assumed to be governed by a uniform distribution, i.e. ${B_i}\sim U(0,2{B_{ave}}),{\rm{ }}1 \le i \le N$. Considering the real bandwidth requirement from $N$ vehicles, the bandwidth requirement ${B_i} \sim U(0,2{B_{ave}}),{\rm{ }}1 \le i \le N$ from $n,{\rm{ }}n \le B/{B_{ave}}$ vehicles is assumed to be less than the average bandwidth requirement ${B_{ave}}$ in the fog cell. The throughput of a vehicle is ${C_j}$ when the bandwidth of a vehicle is allocated by ${B_j}$. For the traditional average bandwidth allocation scheme, the maximum available bandwidth for a vehicle is ${B_{ave}}$ and then the total bandwidth allocated for all vehicles in fog cell is ${B_{tra}} = \sum\limits_{j = 1}^n {{B_j}}  + (N - n) \times {B_{ave}}$. Consequently, the throughput of the fog cell is ${C_{tra}} = \sum\limits_{j = 1}^n {{C_j}}  + (N - n) \times {C_{ave}}$. Based on the proposed adaptive bandwidth allocation scheme, the un-occupied bandwidths of $n$ vehicles can be reused for the other $N - n$ vehicles in the fog cell. The total requirement bandwidth of the other $N - n$ vehicles is $\sum\limits_{j = n + 1}^N {{B_j}} $ and the total available bandwidth of the other $N - n$ vehicles is $B - \sum\limits_{j = 1}^n {{B_j}} $. Therefore, the throughput of a fog cell adopting the adaptive bandwidth allocation scheme is $Min\left\{ {\begin{array}{*{20}{c}}
{\sum\limits_{j = 1}^N {{B_j}} }&,&B
\end{array}} \right\}$.

When the parameter are configured as $C = 1000$ Mbps and ${C_{ave}} = 33$ Mbps, the throughput of a fog cell is compared by two bandwidth allocation schemes in Fig. 5. It is shown that the throughput of a fog cell with the adaptive bandwidth allocation scheme is always larger than the throughput of a fog cell with the average bandwidth allocation scheme. The reason is that the unoccupied bandwidths in the average bandwidth allocation scheme could be utilized in the adaptive bandwidth allocation scheme. When the bandwidth allocation scheme is given, the throughput of a fog cell first increases with the increase of the number of vehicles in a fog cell. When the number of vehicles is larger than 30, the throughput of a fog cell keeps a stationary. The reason is that all available bandwidths in a fog cell have already been allocated for vehicles. In this case, there are not any bandwidths to be allocated for additional vehicles even the number of vehicles is larger than a specified threshold. Consequently, the throughput of a fog cell has to keep a stationary when the number of vehicles is larger than a specified threshold.

\begin{figure}[!t]
\begin{center}
\includegraphics[width=6.5in]{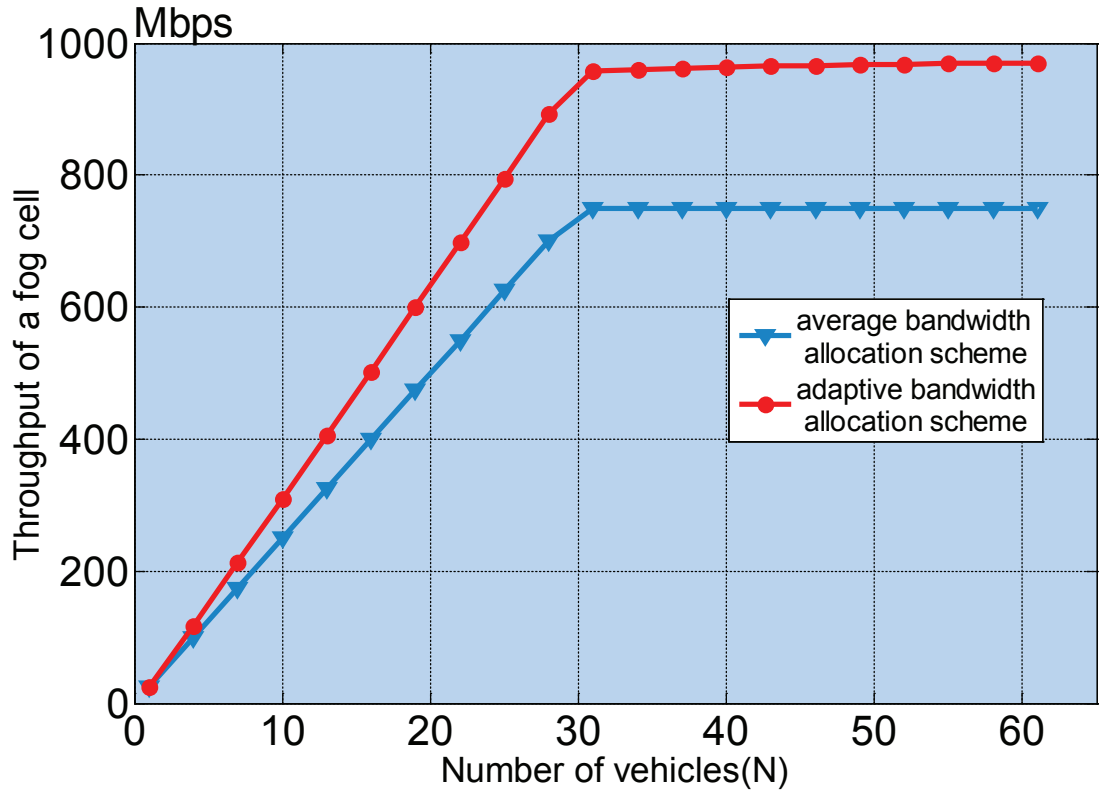}
\caption{Throughput with respect to the number of vehicles in a fog cell}\label{Fig5}
\end{center}
\end{figure}

\section{Challenges of 5G Vehicular Networks}
With the development of 5G mobile communication systems, the high speed wireless communications are satisfied by millimeter wave and massive MIMO technologies. Furthermore, multimedia wireless communications are expected to be realized for 5G vehicular networks. Based on 5G high speed wireless communications, pilotless vehicles are emerging to change our future life. It is well known that future pilotless vehicles need to be supported by high reliable and effective vehicular networks. However, some potential challenges and issues still need to be further investigated for 5G vehicular networks.

\begin{itemize}
  \item
       The low delay issues. When pilotless vehicles are deployed for city transport systems, not only traffic information but also road information should be transmitted to pilotless vehicles by vehicular networks. In general, safety message transmissions have a very low delay constrain, such as less than one millisecond. When there exist many relay vehicles for a multi-hop relay vehicular network, the transmission delay of the warning message will be larger than a given threshold. For some extreme cases, the delay issues would conduce to fatal accidents. How to optimize the route solution is still a key technology for 5G vehicular networks.
  \item
       The frequently handover issues. In this paper the fog cell is proposed to solve the frequently handover between the RSU and vehicles. However, the handover among vehicles is still an issue for the multi-hop relay link in a fog cell. When a lot of vehicles are handed off between adjacent fog cells, the handover will be simultaneously generated for fog cells and the multi-hop relay links. In this case, the complexity of handover is obviously increased for 5G vehicular networks. Moreover, the propagation delay of the warning message needs to be minimized for vehicles handover in 5G vehicular networks.
  \item
       The high service efficiency challenges. For the future pilotless vehicles, vehicles are not only the transport tools but also entertainment centers for users. Different multimedia services need to be provided by 5G vehicular networks. Hence, the massive wireless traffic is expected to increase for 5G vehicular networks. It is a great challenge to improve the service efficiency for 5G vehicular networks.
  \item
       The architecture of 5G vehicular networks. To reduce the transmission delay of the warning messages, the distributed network architecture is adopted for the fog cell of 5G vehicular networks. To support ITSs, the centralized network architecture is adopted for the core network of 5G vehicular networks. In this case, the SDN is proposed to flexibly connect different types of network architectures. However, the scalability and compatibility of 5G vehicular networks are great challenges, especially there exist two types of network architectures in 5G vehicular networks.
\end{itemize}

\section{Conclusions}
With the development of pilotless vehicles, vehicular networks have to face rigorous performance requirements in future ITSs. 5G mobile communications, cloud computing and SDN technologies provide potential solutions for future vehicular networks. In this paper we propose a new architecture of 5G software defined vehicular networks which integrate 5G mobile communications, cloud computing and SDN technologies. Moreover, fog cells are established at the edge of 5G software defined vehicular networks which utilize multi-hop relay networks to reduce the frequent handover between the RSU and vehicles. Simulation results indicate that there exist a minimum transmission delay of 5G software defined vehicular networks considering different vehicle densities. Moreover, the throughput of fog cells in 5G software defined vehicular networks is better than the throughput of traditional transportation management systems. When the proposed challenges of 5G vehicular networks have been solved, 5G software defined vehicular networks could provide enough flexibility and compatibility to satisfy future pilotless vehicles and ITSs.

\begin{IEEEbiographynophoto}{Xiaohu Ge}
 {[}M'09-SM'11{]} (xhge@hust.edu.cn) is currently a full professor
with the School of Electronic Information and Communications at Huazhong
University of Science and Technology (HUST), China and an adjunct
professor with at with the Faculty of Engineering and Information
Technology at University of Technology Sydney (UTS), Australia. He
received his Ph.D. degree in information and communication engineering
from HUST in 2003. He is the director of China International Joint
Research Center of Green Communications and Networking. He has published
more than 140 papers in international journals and conferences. He
served as the general Chair for the 2015 IEEE International Conference
on Green Computing and Communications (IEEE GreenCom). He has served
as an Editor for the \emph{IEEE Transaction on Green Communications
and Networking}, etc.
\end{IEEEbiographynophoto}

\begin{IEEEbiographynophoto}{Zipeng Li}
 received his Bachelor\textquoteright s degree in telecommunication engineering and Master\textquoteright s degree in communication and information system from Huazhong University of Science and Technology, Wuhan, China, in 2011 and 2014, respectively, where he is currently working toward his Doctor \textquoteright s degree. His research interests include vehicular networks and 5G mobile communication systems.
\end{IEEEbiographynophoto}

\begin{IEEEbiographynophoto}{Shikuan Li}
 received his Bachelor\textquoteright s degree in communication and
information system from Huazhong University of Science and Technology,
Wuhan, China, in 2016, where he is currently working toward his Master\textquoteright s
degree. His research interests include vehicular networks and
5G mobile communication systems.
\end{IEEEbiographynophoto}

\end{document}